\documentclass[10pt,reqno]{amsart}
\usepackage{amsmath}
\usepackage{hyperref}

\newcommand{\qbin}[2]{\genfrac{[}{]}{0pt}{}{#1}{#2}}

\newcommand{\cS}{{\mathcal S}}

\newcommand{\cF}{{\mathcal F}}

\newcommand{\A}{{\mathbb A}}
\newcommand{\CC}{{\mathbb C}}

\newcommand{\ZZ}{{\mathbb Z}}

\newcommand{\KK}{{\mathbb K}}
\newcommand{\NN}{{\mathbb N}}

\newcommand{\II}{{\mathbb I}}
\newcommand{\id}{{\mathbf 1}}

\newcommand{\blin}[3]{#1 \cdot #2 \cdot #3}
\newcommand{\sut}{\mathfrak{su}(2)} 
\newcommand{\sud}{\mathfrak{su}(3)}
\newcommand{\ue}{\mathfrak{u}(1)}

\newcommand{\vac}{\lvert 0 \rangle}
\DeclareMathOperator*{\tr}{Tr}

\begin{document}


\title[Parafermion statistics]{Parafermion statistics and the
application to non-abelian quantum Hall states}


\author[E.~Ardonne]{Eddy Ardonne}
\address{Institute for Theoretical Physics, University of Amsterdam,
Valckenierstraat 65, 1018~XE~~Amsterdam, The Netherlands}
\email{ardonne@science.uva.nl}

\thanks{ITFA-01-29}


\begin{abstract}
The (exclusion) statistics of parafermions is used to study degeneracies
of quasiholes over the paired (or in general clustered) quantum Hall states.
Focus is on the $\ZZ_k$ and $\sud_k/\ue^2$ parafermions, which are
used in the description of spin-polarized and spin-singled clustered
quantum Hall states.
\end{abstract}

\keywords{Parafermions, quantum Hall effect, state counting}

\maketitle


\section{Introduction}
\label{intro}

In the last decade, low-dimensional systems in which the fundamental
(quasi) particles do not satisfy bose or fermi statistics have received
a lot of attention. Among the most famous examples are the fractional 
quantum Hall (fqH) systems. The quantum Hall states at simple 
filling fractions $\nu = \frac{1}{M}$, where $M$ is an odd integer,
are understood in terms of the famous Laughlin states \cite{la83a}.
Quasiparticles (or quasiholes) over these states carry fractional
charge and satisfy fractional statistics.

In the last few years, generalizations of these states have been
under investigation. Among these generalizations are so-called
paired (or more general, clustered) quantum Hall states. The notion
of clustering will be explained in the next section. In the
definition of these states, parafermions play a predominant role.
Ultimately, it is the presence of these parafermions which cause
the quasiholes to have peculiar statistics properties, namely,
they obey `non-abelian' statistics (see, for an early reference,
\cite{more91a}).
  
The simplest of the paired quantum Hall states is the pfaffian
state proposed by G.~Moore and N.~Read \cite{more91a}; it is believed
that this states describes the quantum Hall effect at filling fraction
$\nu= \frac{5}{2}$. This quantum Hall effect is special in the sense
that it is the only quantum Hall effect observed at a filling fraction
with an {\em even} denominator (in single layer samples)
\cite{wiei87a,paxi99a}.
For a recent review on this subject, see \cite{re01a}. 
Clustered analogues of the paired pfaffian state were proposed by
N.~Read and E.~Rezayi (RR) \cite{rere99a}. The states mentioned above 
are all spin-polarized; spin-singlet analogues were proposed in
\cite{arsc99a}. Recently, the bosonic versions of the RR states
were shown to be relevant in the context of rotating Bose-Einstein
condensates \cite{cowi01a}.

The statistics properties of the parafermion fields will be 
investigated in this paper, with the intention of obtaining 
closed form expressions for the ground state degeneracy of clustered
quantum Hall states in the presence of quasihole excitations,
as described in \cite{gure00a,arre01a}. This state counting problem
is interesting for the following reasons. The clustered quantum 
Hall states can be seen as ground states of a hamiltonian
with an (ultra local) interaction between the electrons. Finding 
the ground state degeneracy of this hamiltonian can be done in
a conformal field theory (CFT) approach, relying heavily on
the statistics properties of the parafermionic fields.
Another approach is by numerically diagonalizing the interaction
hamiltonian for a small number of electrons. This method can 
serve as a check on the analytical results of the first approach. 
Thus, the quasihole degeneracies of a system of interacting
electrons can be understood in terms of parafermionic statistics!

In the context of the the spin-polarized states of Read and Rezayi,
the $\ZZ_k$ (or $\sut_k/\ue$) parafermions are the relevant parafermions.
For the `non-abelian spin-singlet' (NASS) states of \cite{arsc99a},  
the relevant parafermions are the parafermions related to $\sud_k/\ue^2$
(see \cite{ge87a} for a discussion on general parafermion CFTs). 

The plan of the paper is as follows. We first recall in which way
(clustered) quantum Hall states can be constructed in CFTs
(section \ref{construction}).
In section \ref{numerics} we will shortly indicate the setup of
numerical diagonalization studies, because we need to adapt the
calculations to the setup in which these
studies are done. The general structure of the counting formulas will
be indicated in section \ref{countform}. It will become clear that
the degeneracy consists of an intrinsic and an orbital part, which
need to be combined in
the right way. The intrinsic degeneracy factors need to be split
to make this possible. The remainder of the paper is devoted to this
task, as no explicit expressions for these `split degeneracies'
were known for general level $k>2$ for the states under consideration
in this paper. We will explain the procedure to obtain these expressions
using the $\sud_2/\ue^2$ parafermions of the NASS state at level $k=2$
as an example (see \cite{arre01a}).
The first step is to find a basis for the (chiral) spectrum of the
parafermion CFT. Using this basis, recursion relations for
truncated characters will be derived (section \ref{su32basis}).
These recursion relations can be solved using the results of section
\ref{recrel}, providing expressions for the truncated characters.
From the explicit truncated characters, the `split degeneracies' can
be extracted.
Finally the counting formula for the paired spin-singlet
states is obtained in section \ref{nassk2}, filling in some of the
details of the discussion in \cite{arre01a}. In section \ref{rrcount},
counting formulas for the RR states at general level $k$ are
obtained, while section \ref{nasscount} deals with the counting
formulas for the general $k$ NASS states. Section \ref{discussion} is
devoted to discussions. In this paper, we do not go into the details
of the numerical analysis, but refer to the papers
\cite{rere96a,rere99a,gure00a,arre01a}. For all the cases checked,
the results of the numerical diagonalization studies are exactly
reproduced by the counting formulas. 

\section{Construction of clustered quantum Hall states}
\label{construction}
Clustered quantum Hall states are constructed as correlators in
certain CFTs. But before we come to this construction, let us first
explain the nomenclature `clustered states'. 
Clustered states (with order $k$ clustering) have the following form
\begin{equation}
\Psi^M_k (\{ z_i \} ) = \Phi^k_{\rm bos} (\{ z_i \})
\prod_{i<j} (z_i-z_j)^M \ ,
\end{equation}
where $\Phi_{\rm bos}^k$ is a bosonic factor (symmetric under the exchange
of any two coordinates) with the property
\begin{equation}
\left\{
\begin{array}{l}
\Phi_{\rm bos}^k (z_1 = z_2 = \dots = z_i) \neq 0 \quad i\leq k \\
\Phi_{\rm bos}^k (z_1 = z_2 = \dots = z_i) = 0 \quad i>k \ , \\
\end{array}
\right. 
\end{equation}
i.e. the wave function vanishes as soon as $k+1$ or more particles
are at the same location. Note that we omitted the exponential factors of
the wave functions.   
The factor $\prod_{i<j} (z_i-z_j)^M$ is the familiar Laughlin factor. 
The fermionic quantum Hall states all have $M$ odd, while the bosonic
versions would have $M$ even. Though changing $M$ changes many properties
of the states, such as the filling fraction, the quasihole charge and
statistics, the quasihole degeneracies are unaffected, and the counting
formulas presented in this paper hold for any $M$. Note that this degeneracy
is not to be confused with the `torus degeneracy', which in fact does
depend on $M$.

Following the pioneering paper \cite{more91a}, quantum Hall states can,
under certain restrictions, be defined as correlators in CFTs. 
For the clustered quantum Hall states, CFTs with affine Lie algebra
symmetry are used. The operators creating the electrons in general
consist of a parafermion field and a vertex operator of 
a set of $r$ chiral boson fields ($r$ is the rank of the affine
Lie algebra). These boson fields take care of the charge, spin
and possibly other quantum numbers associated to the particles. 
The electron creation operators take the form
\begin{equation}
V_{\rm el}(z_i) = \psi_\alpha(z_i)
:e^{i \boldsymbol{\beta} \cdot \boldsymbol{\varphi}}(z_i) : \, ,
\end{equation}
where $\psi_\alpha$ is a parafermion field and $\boldsymbol{\varphi}$
is a set of free boson fields. 
The quantum Hall states can be now be defined as follows
\begin{equation}
\Psi (\{ z_i \}) = \lim_{z_\infty \rightarrow \infty } z_\infty^{a}
\langle V_{{\rm el},1} V_{{\rm el},2} \cdots V_{{\rm el},N}
:e^{-i {\bf b} \cdot \boldsymbol{\varphi}}(z_\infty): \rangle \, .
\end{equation}
The parameter ${\bf b}$ in the background charge needs to be chosen in such
a way that the overall correlator is charge neutral. $a$ is chosen such
that the effect of the background charge does not go to zero in the limit
$z_\infty \rightarrow \infty$, whilst keeping the result finite. 
Moreover, the (unique) fusion of the parafermion fields $\psi$ must
result in the identity operator $\id$ in order to get a non-zero
correlator. This in general puts a restriction on the number of electrons
$N$. In this paper, we will not give explicit results for the
correlators (and thereby the wave functions $\Psi_k^M$).
These wave functions and other properties of the ground state, such 
as the filling fraction, can be found in \cite{rere99a} for the Read-Rezayi
states and in \cite{arre01a} for the spin-singlet analogues proposed
in \cite{arsc99a}. A different form of the Read-Rezayi wave functions
can be found in \cite{cage01a}. Also, wave functions for
states with quasiholes (see below) are presented there.  

States with excitations (quasiholes) present are also defined
in terms of correlators in the same CFT as the parent state, by
inserting the corresponding quasihole operator in the correlator.
Certain properties of these quasiholes can be studied via the
corresponding correlators. A constraint on these operators follows
from the condition that the wave function has to be analytical
in the electron coordinates (the lowest Landau level condition).
This implies that the quasihole operators need to be {\em local}
with respect to the electron operators. This constrains the
quasihole operators to be of the following form
\begin{equation}
V_{\rm qp} (w) = \sigma_\varpi
:e^{i \boldsymbol{\beta}' \cdot \boldsymbol{\varphi}}(w): \ ,
\end{equation}
where $\sigma_\varpi$ is a `spin field' of the parafermion CFT. Note that 
the nomenclature spin field does not refer to the electron spin. 
Again, one has to insert a background charge to enforce charge neutrality.
Also, upon fusing the parafermion and spin fields, one has to 
fuse to the identity operator $\id$ in the last step, to obtain a non-zero
correlator. This constrains the possible particles in the correlator.  
The fusion of the spin fields is not unique: in general there
is more than one fusion channel. This implies that a correlator with
several quasihole operators in general stands for more than one 
quantum Hall state. In other words, the clustered quantum Hall states with 
quasiholes are degenerate. It is this degeneracy which lies at the heart
of the non-abelian statistics.

One immediate question one can ask is 
how many states does one describe with such correlators?
In fact, this question can be answered in two independent ways. The 
first one is via a numerical diagonalization study of interacting 
electrons on the sphere, in the presence of a magnetic field. 
In this paper, we will follow the second approach, which is analytical,
and uses the conformal field theory of the underlying parafermions.
But before we come to this point,
we have to spend some words on the numerical approach as well, in
order to be able to adapt the analytical approach to the numerical
setup. This is the subject of the next section; for a more details,
we refer to \cite{arre01a}.

\section{The setup of the numerical studies}
\label{numerics}

Though we will not describe numerical diagonalization studies
in depth in this paper, it is necessary to point out briefly in which
setup they are done, because we need to adapt our calculations
to be able to compare results. 
The numerical diagonalization is most easily done on the
sphere. The interaction between the electrons is chosen such that the 
clustered state under investigation is the unique ground state
(in the absence of quasihole excitations). Note that this interaction
is an ultra local, many-body interaction, rather different from the
long range Coulomb interaction.
To `tune' to the right filling fraction, a specific number of flux
quanta need to penetrate the sphere. States with quasiholes can be
studied by increasing the number of flux quanta (but keeping all the
other parameters the same); this results in the 
creation of quasiholes, as can be seen from the Laughlin
gauge argument. The number of flux quanta needed for a state on
the sphere with quasiholes is given by
\begin{equation}
N_\phi = \frac{1}{\nu} N - \cS + \Delta N_\phi \ ,
\end{equation}
where $N$ is the total number of electrons, and $\Delta N_\phi$ the 
number of excess flux quanta, needed for the creation of the quasiholes.
$\cS$ is an integer constant depending on the state under
investigation. Also, the number of quasiholes which are created by
increasing the number of flux quanta by one depends on the state
under investigation. For the spin-polarized RR states, this relation
is given by $n = k \Delta N_\phi$, where $n$ is the number of quasiholes.
For the spin-singlet analogues, we have
$n=n_\uparrow+n_\downarrow = 2k\Delta N_\phi$.

For the clustered quantum Hall states with quasiholes present, the
ground state is degenerate (for the ultra local interaction).
The degeneracy consists of two parts.
First of all, there is an orbital degeneracy, which is caused by the
fact that in this setup, the quasiholes are non-local. This orbital
degeneracy is not specific for clustered states; it is also present
for the (unpaired) Laughlin states. For a system in which the
quasiholes are localized, this degeneracy would not be present.
Secondly, there is an intrinsic degeneracy, which stems
from the non-trivial fusion rules of the spin fields, needed to create
quasihole excitations. This source of degeneracy is special for
the clustered states. In this paper, we will focus on this intrinsic
degeneracy and obtain analytical expressions, which allow the combination
with the orbital degeneracy factors. This provides us with explicit 
expressions for the degeneracy of the ground states, in the presence of 
quasiholes.

As spin and angular momentum are good quantum numbers, all the states
obtained from the numerical diagonalization fall into spin and angular
momentum multiplets. The structure of the counting formulas is such that
also the multiplet structure can be extracted. 

\section{Degeneracy factors and counting formulas}
\label{countform}
The intrinsic degeneracy is caused by the non-trivial fusion rules of the
spin fields. As an example, we will use the spin fields of the 
$\sud_2/\ue^2$ parafermionic CFT. The fields and their 
fusion rules in this theory can be determined according to the 
methods of \cite{ge87a} and are summarized in table \ref{fusrul}.
We use the notation introduced in \cite{arre01a}. The parafermion fields
are denoted by $\psi$, and all have conformal dimension
$\Delta_\psi = \tfrac{1}{2}$. In particular, 
$\psi_1$,$\psi_2$,$\psi_{12}$ correspond to the roots $\alpha_1$,$-\alpha_2$
and $\alpha_1+\alpha_2$ of $su(3)$, respectively.
The spin fields
$\sigma_\uparrow$,$\sigma_\downarrow$,$\sigma_3$ and $\rho$
are related to the weights of $su(3)$ and their conformal dimensions
are given by $\Delta_\sigma = \tfrac{1}{10}$ and $\Delta_\rho = \tfrac{3}{5}$.
\begin{table}
\begin{tabular}{c|c|c|c|c|c|c|c}
$\times$ & $\sigma_\uparrow$ & $\sigma_\downarrow$ &
$\sigma_3$ & $\rho$ & $\psi_1$ & $\psi_2$ & $\psi_{12}$
\\ \hline
$\sigma_\uparrow$ & $\id + \rho$ &&&&&&  \\

$\sigma_\downarrow$ & $\psi_{12}+\sigma_3$ & $\id+\rho$ &&&&&\\

$\sigma_3$ & $\psi_1 + \sigma_\downarrow$ &
$\psi_2 + \sigma_\uparrow$ & $\id + \rho$ &&&&\\

$\rho$ & $\psi_2$ + $\sigma_\uparrow$ &
$\psi_1 + \sigma_\downarrow$ & $\psi_{12} + \sigma_3$ &
$\id + \rho$ &&&\\

$\psi_1$ & $\sigma_3$ & $\rho$ & $\sigma_\uparrow$ &
$\sigma_\downarrow$ & $\id$ &&\\

$\psi_2$ & $\rho$ & $\sigma_3$ & $\sigma_\downarrow$ &
$\sigma_\uparrow$ & $\psi_{12}$ & $\id$ &\\

$\psi_{12} $ & $ \sigma_\downarrow$ & $\sigma_\uparrow$ &
$\rho$ & $\sigma_3$ & $\psi_2$ & $\psi_1$ & $\id$ \\
\end{tabular}
\vskip 2mm \caption{Fusion rules of the parafermion and spin
fields associated to the parafermion theory $\sud_2/\ue^2$
introduced by Gepner \protect\cite{ge87a}.} \label{fusrul}
\end{table}
The fusion of an arbitrary number of $\sigma_{\uparrow,\downarrow}$
fields can be depicted in a Bratteli diagram (see also
\cite{arre01a}).
\begin{figure}[h]
\setlength{\unitlength}{1mm}
\begin{picture}(67,22)(-10,-3)
\put(0,0){\vector(1,1){5}}
\put(5,5){\vector(1,1){5}}
\put(5,5){\vector(1,-1){5}}
\put(10,0){\vector(1,1){5}}
\put(10,10){\vector(1,1){5}}
\put(10,10){\vector(1,-1){5}}
\put(15,5){\vector(1,1){5}}
\put(15,5){\vector(1,-1){5}}
\put(15,15){\vector(1,-1){5}}
\put(20,0){\vector(1,1){5}}
\put(20,10){\vector(1,1){5}}
\put(20,10){\vector(1,-1){5}}
\put(25,5){\vector(1,1){5}}
\put(25,5){\vector(1,-1){5}}
\put(25,15){\vector(1,-1){5}}
\put(30,0){\vector(1,1){5}}
\put(30,10){\vector(1,1){5}}
\put(30,10){\vector(1,-1){5}}
\put(34.5,0){\ldots}
\put(34.5,10){\ldots}
\put(40,5){\vector(1,1){5}}
\put(40,5){\vector(1,-1){5}}
\put(40,15){\vector(1,-1){5}}
\put(45,0){\vector(1,1){5}}
\put(45,10){\vector(1,-1){5}}
\put(50,5){\vector(1,-1){5}}
\put(0,-3){0}
\put(10,-3){2}
\put(20,-3){4}
\put(30,-3){6}
\put(45,-3){$n$-2}
\put(55,-3){$n$}
\put(-10,0){$\id$,$\psi_{12}$}
\put(-10,5){$\sigma_\uparrow$,$\sigma_\downarrow$}
\put(-10,10){$\rho$,$\sigma_3$}
\put(-10,15){$\psi_1$,$\psi_2$}
\put(56,0){$\id$}
\put(21,8.1){*}
\end{picture}\\
\caption{Bratteli diagram for the spin fields of
$\sud_2/\ue^2$.}
\label{brat}
\end{figure}
Each arrow in the diagram in figure \ref{brat} stands for either
a $\sigma_\uparrow$ or $\sigma_\downarrow$ field. The arrow starts
at a certain field which can only be one of the fields on the left
of the diagram at the same height. This last field is fused with
the one corresponding to the arrow, while the arrow points at a field
present in this fusion. As an example, the arrows starting at the
$*$ are encoding the fusion rules
$\rho \times \sigma_{\uparrow
(\downarrow)} = \psi_{2 (1)} + \sigma_{\uparrow (\downarrow)}$
and
$\sigma_3 \times \sigma_{\uparrow (\downarrow)} = \psi_{1 (2)} +
\sigma_{\downarrow (\uparrow)}$.
One checks that the diagram is in
accordance with the first two columns of Table \ref{fusrul}.

From figure \ref{brat}, one immediately reads off that in general
there is more than one fusion path of spin fields with leads to the
identity (possibly the identity is reached only after the fusion 
with the parafermion fields $\psi_{1,2}$ of the electron operators). 
It is easily seen that the number of fusion channels
starting from and terminating at $\id$ while $n_\uparrow$
$\sigma_\uparrow$ and $n_\downarrow$ $\sigma_\downarrow$ 
spin fields are fused is given by
\begin{equation} 
d_{n_\uparrow,n_\downarrow} = \cF (n_\uparrow+n_\downarrow -2) \, ,
\end{equation}
where $\cF (n)$ is the $n$-th Fibonacci number, defined by
$\cF(n) = \cF(n-1)+\cF(n-2)$ with the initial conditions
$\cF(0)=1$ and $\cF(1)=1$.
Next to this intrinsic degeneracy, there is an orbital degeneracy. 
These orbital degeneracy factors can be found in \cite{gure00a,arre01a}
for the states discussed in this paper. These factors have the
general form
\begin{equation}
\prod_i \binom{\frac{n_i-F_i}{k} + n_i}{n_i} \ .
\end{equation}
The product is over the types of quasiholes, while 
the numbers $F_i$ are interpreted as the number of `unclustered'
particles in the state. In the correlators, these correspond to
the fundamental parafermions $\psi_i$. For each fusion path, these
numbers can be different, implying that we have to split the
intrinsic degeneracy according to these numbers. We denote
these `split degeneracy factors' by $\{ \}_k$. Explicitly, we
have $\{ \begin{smallmatrix} n\\F \end{smallmatrix} \}_k$ and
$\{ \begin{smallmatrix} n_\uparrow & n_\downarrow \\
F_1 & F_2 \\ \end{smallmatrix} \}_k$ for the RR and NASS states,
respectively.

Using the notation above, the counting formula for the clustered
spin-singlet quantum Hall states take the following form
\begin{equation}
\label{genfor}
\#_{\rm NASS} (N,\Delta N_\phi,k) = 
\sideset{}{'}\sum_{N_{\uparrow,\downarrow},n_{\uparrow,\downarrow},F_{1,2}}
\begin{Bmatrix}
n_{\uparrow} & n_{\downarrow} \\
F_{1} & F_{2} \\
\end{Bmatrix}_k
\binom{\frac{N_{\uparrow}-F_{1}}{k} + n_{\uparrow}}{n_\uparrow}
\binom{\frac{N_{\downarrow}-F_{2}}{k} + n_{\downarrow}}{n_\downarrow}
\ ,
\end{equation}
where the prime on the summation indicates the presence of
constraints (see below eq. \eqref{ncgk}). The equivalent counting
formula for the Read-Rezayi states is given in eq. \eqref{rcgk}.

The new result of this
paper are the explicit split degeneracy factors for the $\ZZ_k$
and $\sud_k/\ue^2$ parafermions at level $k>2$ 
(for $k=2$, these factors can be found in \cite{rere96a} and
\cite{arre01a} respectively). Previously, these factors
for the $\ZZ_k$ parafermions ($k>2$) could only be characterized
via recursion relations, see \cite{gure00a,bosc99a}.
Note that the results in this paper are easily extended
to the more general $\mathfrak{su}(N)_k/\ue^{N-1}$ parafermions.

We will now briefly outline in which way the split degeneracy factors
are obtained. The starting point is the character of the parafermionic
CFT. The symbols $\{ \}_k$ can be extracted from finitized forms of these
characters \cite{gure00a} (see also \cite{sc97a}).
Recursion relations for these finitized characters can be derived from 
an explicit basis of the parafermionic CFT. These recursion relations
will be written in a way that allows for an explicit solution,
from which the symbols $\{ \}_k$ can be extracted. In the sections
\ref{su32basis}-\ref{nassk2}, we will demonstrate this for the
level $k=2$ spin-singlet states of \cite{arsc99a}.

\section{A basis for the $\sud_2/\ue^2$ parafermion theory}
\label{su32basis}

In this section, we briefly describe how an explicit
basis for the chiral spectrum of the $\sud_2/\ue^2$ parafermion 
CFT is formed. 
The starting point is the chiral character for the parafermions
in the $\sud_2/\ue^2$ conformal field theory. 
This character can be written in the form of a
`Universal Chiral Partition Function' (UCPF) see, for instance,
\cite{bemc99a,arbo01b}. This character reads \cite{kuna93a}
\begin{equation} 
\label{nasschar}
{\rm ch}(x_1,x_2;q,k=2) 
= \sum_{n_1,n_2} \frac{q^{(n_1^2+n_2^2-n_1 n_2)/2}}
{(q)_{n_1}(q)_{n_2}} x_1^{n_1} x_2^{n_2} \ .
\end{equation}
In this character, $x_i = e^{\beta \mu_i}$ are  fugacities of the
particles, and $q = e^{\beta \varepsilon}$ ($\beta$ is the
inverse temperature). $(q)_a$ is defined by
$(q)_{a}= \prod_{k=1}^a (1-q^k)$ for $a > 0$ and $(q)_0 =1$.

The bilinear form in the exponent of $q$ is described by the matrix
\begin{equation}
\label{kmat}
\KK =  \begin{pmatrix}
1 & -\frac{1}{2} \\ -\frac{1}{2} & 1 \\ \end{pmatrix}  \ .
\end{equation}
The same matrix also describes the exclusion statistics of these 
parafermions. More information on the relation between exclusion
statistics and the UCPF can be found in \cite{arbo01a}.

A basis for a CFT can be thought of as a set of states spanning the
chiral Hilbert space. This set of states can be written as 
a (set of) vacuum state(s), on which creation operators act.
The parafermions $\psi_{1,2} (z)$ in the $\sud_2/\ue^2$ theory can
be expanded in modes as
\begin{equation}
\label{modexp}
\psi_{1,2} (z) = \sum_{m\in\ZZ} z^{-m} \psi_{m-\frac{1}{2}}^{1,2} \ .
\end{equation}
As usual, the modes $\psi_m$ with negative index are the creation
operators while the modes with positive index annihilate the vacuum
\begin{equation}
\psi_m \vac = 0 \quad m > 0 \ .
\end{equation}
The set of states
\begin{equation}
\psi^{a_n}_{-s_n} \psi^{a_{n-1}}_{-s_{n-1}} \cdots \psi^{a_1}_{-s_1}
\vac
\end{equation} 
is overcomplete, because of the (generalized) commutation rules of the
parafermions. In the following, we will point out which restrictions
on the indices $s_i$ will remove the `overcompleteness'. 
In doing so, we will follow the exclusion interpretation of the
K-matrix as closely as possible and concentrate on the lowest
possible `energy' (given by $L_0=\sum_i s_i$) for a certain number of 
applied fields first. The ordering of the modes $\psi^{1,2}$ is
such that we apply the $\psi^1$ modes first. 
From \eqref{modexp} it follows that the simplest non-trivial state
is
\begin{equation}
\psi^1_{-1/2} \vac \ .
\end{equation} 
Interpreting the matrix \eqref{kmat} as the exclusion statistics matrix,
the minimal spacing between two $\psi^1$ modes is $1$, thus the
state with two $\psi^1$'s acting on the vacuum with minimal energy is
\begin{equation}
\psi^1_{-3/2} \psi^1_{-1/2} \vac \ .
\end{equation} 
The extension to $n_1$  $\psi^1$ modes is simple
\begin{equation}
\psi^1_{-(2n_1-1)/2} \cdots \psi^1_{-3/2} \psi^1_{-1/2} \vac \ .
\end{equation}
Note that if this were the whole story, we would describe the (free)
Majorana fermion. 
The spacing between $\psi^2$ modes is the same as for the $\psi^1$
modes. However,
if one acts with $\psi^2$ on a state in which $\psi^1$ modes
are already present, one has to take into account the mutual statistics
between $\psi^1$ and $\psi^2$ modes, which is $-1/2$, according to
\eqref{kmat}. Thus the energies of the $\psi^2$ modes have an extra shift
of $-n_1/2$, resulting in the following states (with minimal energy)
\begin{equation}
\psi^2_{-(2n_2-1-n_1)/2} \cdots \psi^2_{-(3-n_1)/2} \psi^2_{-(1-n_1)/2}
\psi^1_{-(2n_1-1)/2} \cdots \psi^1_{-3/2} \psi^1_{-1/2} \vac \ .
\end{equation}
The (dimensionless) energy associated to this state is
$\tfrac{n_1^2 + n_2^2 - n_1 n_2}{2}$, precisely the exponent
of $q$ in the character \eqref{nasschar}.
To obtain all the possible states, one has to allow states with higher
energies as well. As usual \cite{bosc99a}, the energies of all the modes
can have integer shifts, under the restriction that modes acting on
a state have larger energies than the modes of the same type which 
have been applied earlier. This results in the following set of states
\begin{equation}
\label{nassbasis}
\psi^2_{-(2n_2-1-n_1)/2-t_{n_2}} \cdots \psi^2_{-(3-n_1)/2-t_2}
\psi^2_{-(1-n_1)/2-t_1}
\psi^1_{-(2n_1-1)/2-s_{n_1}} \cdots \psi^1_{-3/2-s_2}
\psi^1_{-1/2-s_1} \vac \ ,
\end{equation}
with $s_{n_1} \geq \ldots \geq s_2 \geq s_1 \geq 0$ and
$t_{n_2} \geq \ldots \geq t_2 \geq t_1 \geq 0$ ($s_i , t_j \in \NN$).

Up to now, we used the special ordering of applying modes to
the vacuum, namely, all the $\psi^1$ modes first.
This is in fact enough to span the whole chiral spectrum, as can
be seen if we perform the trace over all basis states. More or
less by construction, we obtain the character \eqref{nasschar}. 
However, we also can allow a general ordering of the modes. 
As an example, we take the following state
\begin{equation}
\label{order1}
\psi^2_{-0} \psi^1_{-1/2} \vac \ .
\end{equation}
The energy of the $\psi^2$ mode is zero because it gets an extra shift
of $-1/2$ due to the presence of the $\psi^1$ mode.
In spanning the whole chiral spectrum, we can also choose to use
the state, with the order of the modes changed
\begin{equation}
\label{order2}
\psi^1_{-0} \psi^2_{-1/2} \vac \ .
\end{equation}
In this case, the $\psi^1$ mode gets an extra shift of $-1/2$, because
of the presence of the $\psi^2$ mode. Thus, the $L_0$ value is the
same for both states. 
In general, changing the order of two neighbouring $\psi^1$ and
$\psi^2$ modes does not change the $L_0$ value if the extra shifts
are changed in the appropriate way. The extra shift of a field is 
given by $-1/2$ times the number of preceding modes of the other
type. In general, two states related by a reordering of modes are
different, but we can use either of them (but not both) to span the
chiral spectrum. Note that the rules of the spacing between the various
fields is in accordance with the (exclusion) statistics interpretation
of the matrix $\KK$. The character \eqref{nasschar} is obtained by
taking the trace over all the states in the basis \eqref{nassbasis}
\begin{equation}
{\rm ch} (x_1,x_2;q) = \tr x_1^{n_1} x_2^{n_2} q^{L_0} \ .
\end{equation}
We can now define the finitized characters needed in the derivation 
of the symbols $\{ \}_k$ by using the basis described above.
These finitized characters are polynomials which will be denoted
by $Y_{(l,m)}$. These polynomials are traces over the basis
\eqref{nassbasis}, but restricted to the states in which the energy
of the modes of the $\psi_1$ ($\psi_2$) fields are smaller or equal
to $l$ ($m$). Though the total energy
of a state does not depend on the ordering of the modes, the
energies of the individual modes do depend on the ordering, as can be
seen by comparing the states \eqref{order1} and \eqref{order2}.
By restricting the trace over states in which the labels of the 
modes are bounded, we must include a state if there is at least one
ordering in which all the modes satisfy the bounds
imposed. Note that there may be other orderings, in which these
bounds are not satisfied. We write the finitized characters as
\begin{equation}
Y_{(l,m)} (x_1,x_2;q) = \sideset{}{'} \tr_{\leq l , \leq m}
x_1^{n_1} x_2^{n_2} q^{L_0} \ .
\end{equation}
The prime on the trace denotes an important restriction on the
number of modes (denoted by $n_1$ and $n_2$) present in the states.
These numbers must satisfy $n_1 = 2l \pmod 2$ and $n_2 = 2m \pmod 2$. 
This restriction takes into account that
after fusing the spin fields, one ends up in the right sector,
which can be $\id,\psi_1,\psi_2$ or $\psi_{12}$ depending
on the number of spin-up and down electrons. 
This is necessary, because after fusing the spin fields and the
parafermion fields of the electron operators, on has to end with
the identity $\id$, to obtain a non-zero correlator.

The finitized characters $Y_{(l,m)}$ can be written in terms of recursion
relations of the following form
\begin{eqnarray} 
\nonumber
Y_{(l,m)} & = & Y_{(l-1,m)} +
x_1 q^{l-\frac{1}{2}}
Y_{(l-1,m+\frac{1}{2})} \ , \\ \label{lmrec}
Y_{(l,m)} & = & Y_{(l,m-1)} +
x_2 q^{m-\frac{1}{2}}
Y_{(l+\frac{1}{2},m-1)} \ .
\end{eqnarray}
Note that the recursion relations above are
stated in terms of the energy labels of the modes.
The aim we have is finding the number of possible states 
when a certain number of extra flux is added. We therefore
need to make a change to labels which depend on the additional
flux. In fact, we will use the number of particles (given by
$n_\uparrow$ and $n_\downarrow$ in this case) created by
this flux as labels for the finitized partition functions.
Explicitly, we have $l=\tfrac{n_\uparrow}{2}$ and
$m=\tfrac{n_\downarrow}{2}$.
In terms of the number of created quasiholes, the recursion relations
become (compare \cite{arre01a})
\begin{align}
\nonumber
Y_{(n_\uparrow,n_\downarrow)} & = Y_{(n_\uparrow-2,n_\downarrow)} +
x_1 q^{\frac{n_\uparrow-1}{2}}
Y_{(n_\uparrow-2,n_\downarrow+1)} \ , \\
\label{recursion}
Y_{(n_\uparrow,n_\downarrow)} & = Y_{(n_\uparrow,n_\downarrow-2)} +
x_2 q^{\frac{n_\downarrow-1}{2}}
Y_{(n_\uparrow+1,n_\downarrow-2)} \ .
\end{align}
The initial conditions for these recursion relations look as follows
\begin{align}
Y_{(1,0)}&=Y_{(0,1)}=0 \ ,
\nonumber\\
Y_{(0,0)}&=Y_{(2,0)}=Y_{(0,2)}=1 \ ,
\nonumber\\ \label{init}
Y_{(1,1)}&=q^{\frac{1}{2}} x_1 x_2 \  .
\end{align}
The finitized characters are completely described by
\eqref{recursion} and \eqref{init}. In the next section, we will
solve these recursion relations and thereby provide explicit
expressions for the finitized characters.

\section{Recursion relations and solutions} 
\label{recrel}

The recursion relations of the previous section can be solved
explicitly; we will follow the approach of \cite{bo00a}.
The key observation is that the recursion relations
can be matched to general recursion relations, which are
solved in terms of finitizations of universal chiral partition
functions. Consider the following polynomials $P$
\begin{equation} \label{ucpf}
P_{\bf L} ({\bf z};q) = 
\sum_{\bf m} \Bigl( \prod_i z_i^{m_i} \Bigr)
q^{ \frac{1}{2} {\bf m} \cdot \KK \cdot {\bf m} +
{\bf Q} \cdot {\bf m}} \prod_i
\qbin{\bigl({\bf L} + (\II -\KK) \cdot {\bf m} + {\bf u} \bigr)_i}{m_i}
\ .
\end{equation}
In this equation, $\II$ is the identity matrix, $\KK$ the statistics 
matrix and $\bigl[ \begin{smallmatrix} a \\ b \end{smallmatrix}\bigr]$
the $q$-deformed binomial ($q$-binomial)
\begin{equation}
\qbin{a}{b} = 
\left\{
\begin{array}{cc}
\frac{(q)_a}{(q)_b (q)_{a-b}} &
a,b \in \NN \, ; \, b \leq a \\
0 & {\rm otherwise} \\
\end{array}
\right. \ .
\end{equation}
Note that we defined the $q$-binomial to be non-zero only
if both entries are integers greater or equal to zero, 
to avoid additional constraints on the sums in the counting formulas.

From the definition of the $q$-binomials, the following
identity is easily derived
\begin{eqnarray} \label{qident}
\qbin{a}{b} &=& \qbin{a-1}{b} +q^{a-b} \qbin{a-1}{b-1} \ . 
\end{eqnarray}
Replacing the $i$'th $q$-binomial 
factor in \eqref{ucpf} by the right hand side of \eqref{qident},
one finds the following recursion relations
\begin{equation} \label{ucpfrec}
P_{\bf L} ({\bf z};q) = P_{{\bf L}-{\bf e}_i}({\bf z};q) +
z_i q^{-\frac{1}{2} \KK_{ii} + {\bf Q}_i + {\bf u}_i + {\bf L}_i}
P_{{\bf L}-\KK \cdot{\bf e}_i}({\bf z};q) \ .
\end{equation}
The vector ${\bf e}_i$ represents a unit vector in the $i$'th
direction.
We will use the equivalence between \eqref{ucpf} and \eqref{ucpfrec}
frequently, because the recursion relations we encounter in
deriving the counting formulas are all of type \eqref{ucpfrec}.
Of course, upon deriving polynomials from recursion
relations, one has to take the initial conditions into
account. For the counting we need to know the finitizations
of the character formulas, and these can be written in the
form \eqref{ucpf}. Thus, when we solve recursion relations by
polynomials of the form \eqref{ucpf}, the proper initial conditions
are automatically taken into account.

We start by applying the above to the recursion relations
\eqref{lmrec}, resulting in the following expressions for the
truncated characters $Y_{(n_\uparrow,n_\downarrow)}$ 
\begin{equation} \label{expol}
Y_{(n_\uparrow,n_\downarrow)}(x_1,x_2;q) =
\sum_{a,b}
q^{(a^2+b^2-ab)/2} x_1^a x_2^b
\qbin{\frac{n_\uparrow+b}{2}}{a}
\qbin{\frac{n_\downarrow+a}{2}}{b} \ .
\end{equation}
This result will be needed for the final counting formula,
which we give in the next section.

\section{A counting formula for the NASS state at $k=2$}
\label{nassk2}

From the truncated characters of the previous section, we can
obtain the symbols $\{ \}_2$, needed in the counting formula
eq. \eqref{genfor}.
In fact, the symbols $\{ \}_2$ are obtained by taking the limit
$q \to 1$ of the coefficient of $x_1^{F_1} x_2^{F_2}$ in 
eq. \eqref{expol} (see, for instance, \cite{gure00a,arre01a})
\begin{equation}
Y_{(n_\uparrow,n_\downarrow)}(x_1,x_2;1) =
\sum_{F_1,F_2}
x_1^{F_1} x_2^{F_2}
\begin{Bmatrix}
n_\uparrow & n_\downarrow \\
F_1 & F_2 \\
\end{Bmatrix} \, .
\end{equation}
In this limit, the $q$-binomials in \eqref{expol} become
`ordinary' binomials and we find
\begin{equation}
\label{nk2sym}
\begin{Bmatrix}
n_{\uparrow} & n_{\downarrow} \\
F_{1} & F_{2} \\
\end{Bmatrix}_2 = 
\binom{\frac{n_\uparrow+F_2}{2}}{F_1}
\binom{\frac{n_\downarrow+F_1}{2}}{F_2} \ .
\end{equation}
The fact that the finitized characters indeed provide the symbols
$\{ \}$ is rather non-trivial. This connection was first proposed
in \cite{gure00a}.
Some (restricted) `solid on solid' (SOS) models (see, for instance
\cite{anba84a}) can be mapped to the Bratteli diagrams of the spin
fields of the quasiholes. Recursion relations for the partition
functions for these models (at finite size) are in general equivalent
to recursion relations for finitized characters in certain CFTs.
In the case at hand, the corresponding CFT is the parafermion CFT.
This provides a link between the Bratteli diagrams and the parafermion
theories. As a check, on can calculate the number of fusion
paths for the spin fields by summing over the symbols $\{ \}$
and compare to the result obtained from the diagram itself.
In this specific case, the equivalence follows from the structure of
the recursion relations (see for instance \cite{arre01a}),
giving rise to the identity
\begin{equation}
\sideset{}{'}\sum_{F_1,F_2}
\binom{\frac{n_\uparrow+F_2}{2}}{F_1} 
\binom{\frac{n_\downarrow+F_1}{2}}{F_2} 
= \cF (n_\uparrow+n_\downarrow -2) \ .
\end{equation} 
The prime on the summation denotes the constraints
$F_1 \equiv n_\downarrow \pmod 2$ and
$F_2 \equiv n_\uparrow \pmod 2$.
At the level of the wave functions, the degeneracy is due to the 
presence of particles which do not belong to a cluster any more.
At the level of correlators, these unclustered particles correspond
to parafermions $\psi^1$ and $\psi^2$, which act as `cluster
breakers'. In the case of the pfaffian state, this was made explicit
in \cite{rere96a}. 

The counting formula for the NASS state at $k=2$ is obtained
by inserting the symbol \eqref{nk2sym} in the general counting
formula \eqref{genfor}
\begin{equation} \label{nck2}
\# (N,\Delta N_\phi,k=2) =
\sideset{}{'}
\sum_{N_{\uparrow,\downarrow},n_{\uparrow,\downarrow},F_{1,2}}
\binom{\frac{n_\uparrow+F_2}{2}}{F_1} 
\binom{\frac{n_\downarrow+F_1}{2}}{F_2} 
\binom{\frac{N_{\uparrow}-F_{1}}{2} + n_{\uparrow}}{n_{\uparrow}}
\binom{\frac{N_{\downarrow}-F_{2}}{2} + n_{\downarrow}}{n_{\downarrow}}
\ ,
\end{equation}
where the prime on the sum indicates the constraints
$N_\uparrow+N_\downarrow = N$,
$n_\uparrow+n_\downarrow = 4 \Delta N_\phi$ and 
$N_\uparrow-N_\downarrow =  n_\downarrow-n_\uparrow$.

We will now comment on the spin and angular momentum
multiplet structure. As an example, we will write out the
polynomial $Y_{(7,1)}$
\begin{equation}
\label{trupolex}
Y_{(7,1)} =
(q^{\frac{1}{2}}+q^{\frac{3}{2}}+q^{\frac{5}{2}}+q^{\frac{7}{2}}) x_1x_2 +
(q^{\frac{7}{2}}+2q^{\frac{9}{2}}+2q^{\frac{11}{2}}+2q^{\frac{13}{2}}
+q^{\frac{15}{2}}) x_1^3x_2 + q^{\frac{19}{2}} x_1^5 x_2^3 \ .
\end{equation}
After multiplying the coefficient of $x_1^{F_1} x_2^{F_2}$ with 
(in general) $q^{-(n_\uparrow F_1+n_\downarrow F_2)/4}$, one obtains
a sum of terms of the form $q^{l_z}$, which together form a collection of
angular momentum multiplets with quantum numbers $l_z$. For instance,
the coefficient of $x_1^3x_2$ in eq. \eqref{trupolex} encodes two
multiplets, namely $L=2$ and $L=1$.

An alternative way to obtain these results is by associating
angular momentum multiplets to the binomials in eq. \eqref{nck2}.
The binomials $\tbinom{a}{f}$ forming the symbols $\{ \}_2$ need to
be interpreted as the number of ways on can put $f$ fermions in $a$
boxes, which are labeled with
$l_z = -\tfrac{(a-1)}{2}, -\tfrac{(a-1)}{2}+1 , \ldots , \tfrac{(a-1)}{2}$
angular momentum quantum numbers. Each way of putting the $f$ fermions in
$a$ boxes has an $l_z^{\rm tot}$ associated with it. Together, these
$l_z^{\rm tot}$ quantum numbers fall into angular momentum multiplets.
In this way, angular momentum multiplets can be associated with the
binomials. The angular momentum multiplets of the various binomials in
the counting formula need to be added in the usual way.

Though the parafermion theory does not have a proper
$SU(2)$ spin symmetry, one can associate spin quantum
numbers to every state by taking
$S_z = \tfrac{N_\uparrow-N_\downarrow}{2}$. Combining the
spin and angular momentum, one finds that all the states fall
into spin and angular momentum multiplets. 

The numerical diagonalization studies for the NASS state at level
$k=2$ are described in \cite{arre01a}. It is very gratifying to
see that the counting formula eq. \eqref{nck2} does in fact exactly
reproduce the quasihole degeneracies, as well as the multiplet
structure. 

In order to find the counting results for the spin-singlet
states at general level-$k$, 
we first take a closer look at the counting of the 
Read-Rezayi states, which was in fact done in \cite{gure00a}.
Those results however, were stated 
in terms of recursion relations which are difficult
to solve. The advantage of the recursion relations presented
in the next section is that they can easily be solved in terms of
($q$-deformed) binomials, and thus provide explicit expressions
for the symbols $\{ \}_k$.

\section{Counting formulas for the Read-Rezayi states}
\label{rrcount}
The derivation of the counting formulas for the RR states goes along
the same lines as the derivation for the NASS k=2 states as explained
in the previous sections. Therefore, we do not go into full detail,
but concentrate on the parts which need more explanation.

We start with the character of the $\sut_k/\ue$ parafermionic
theory (see \cite{zafa85a}), which can be obtained from
\cite{lepr85a,kuna93a,ge95a}
\begin{equation} \label{chsu2k}
{\rm ch} (x;q,k) =  \sum_{a_i}
\frac{q^{\frac{1}{2}({\bf a}\cdot\CC_{k-1}\cdot{\bf a})}}
{\prod_i (q)_{a_i}} x^{i a_i} \ ,
\end{equation}
where ${\bf a}=(a_1,\ldots,a_{k-1})$ and
$\CC_{k-1} = 2 \A^{-1}_{k-1}$, $\A_{k-1}$ 
being the Cartan matrix of $su(k)$. 
In components, these matrices are given by
\begin{align}
(\A_{k-1})_{i,j} &= 2 \delta_{i,j}-\delta_{|i-j|,1} \nonumber \\
(\A_{k-1}^{-1})_{i,j} &= {\rm min} (i,j)-\frac{i j}{k}
\, \hspace{5 mm} \, i,j = 1,\ldots,k-1  \ .
\end{align}
In fact, $\CC_{k-1}$ is the K-matrix for the $\ZZ_k$ parafermions,
see \cite{arbo01b}. The parafermions in this theory are
$\psi_0,\psi_1,\dots\psi_{k-1}$ ($\psi_0$ is the identity $\id$ and
the labels are defined modulo $k$). 
By writing $x^{i a_i}$ in the character \eqref{chsu2k}, we
take care of the fact that the fugacity of species $i$ is
$i$ times the fugacity of the first type of particle. 
In fact, the $i$'th species can be thought of as a `composite' of
$i$ particles of species $1$. This point of view is supported by 
the fusion rules for these parafermions
$\psi_1 \times \psi_p = \psi_{p+1}$, with $p=1,\ldots,k-1$.
This structure is also present in the K-matrix structure
describing the Read-Rezayi states \cite{gusc99a,arbo00a,arbo01a}.

A basis for the chiral spectrum can be constructed 
in the same way as described in section \ref{su32basis}. 
The shifts in modes between the various fields
are given by the elements of the matrix $\KK$. We will now
proceed by directly giving the corresponding recursion relations
\begin{equation}
Y_{\bf l} (x;q,k)= Y_{{\bf l}-{\bf e}_i} + x^i q^{l_i-\frac{i(k-i)}{k}}
Y_{{\bf l}-\CC_{k-1}\cdot{\bf e}_i} \ .
\end{equation}
The factor $\frac{i(k-i)}{k}$ is the conformal dimension 
of the $i$'th parafermion in the $\ZZ_k$-parafermion theory.
These recursion relations are solved by the following  
polynomials
\begin{equation}
Y_{\bf l} (x;q,k)= \sum_{a_i}
q^{\frac{1}{2}(\blin{\bf a}{\CC_{k-1}}{\bf a})}
\prod_{i=1}^{k-1} x^{i a_i}
\qbin{\bigl({\bf l}+(\II_{k-1}-\CC_{k-1}) \cdot {\bf a}\bigr)_i}{a_i}
\ ,
\end{equation}
where $\II_{k-1}$ denotes the $(k-1)$-dimensional unit matrix.
To obtain the counting results, we have to specify the truncation
parameters $l_i$. As in the NASS case for $k=2$, we will
do this in terms of the number of particles created by the extra flux,
given by $n = k \Delta N_\phi$ for the states under consideration.
Because the chemical potential of species $i$ is $i$ times the chemical
potential of species $1$, the truncation parameter $l_i$ has to 
be scaled with a factor $i$ with respect to $l_1$ (see, for instance
\cite{arbo01a}), which is found to be $l_1=\frac{n}{k}$. This leads to
the following truncation parameters $l_i = \frac{i n}{k}$,
and the truncated characters needed for the counting become
\begin{equation} \label{spfin}
Y_n (x;q,k)= \sum_{a_i}
q^{\frac{1}{2}(\blin{\bf a}{\CC_{k-1}}{\bf a})}
\prod_{i=1}^{k-1} x^{i a_i} 
\qbin{\frac{i n}{k}+\bigl((\II_{k-1}-\CC_{k-1})\cdot {\bf a}\bigr)_i}{a_i}
\ .
\end{equation}
To obtain the symbols
$\{\begin{smallmatrix}n\\F\\\end{smallmatrix}\}_k$ which are needed
for the counting, one has to take the limit $q \to 1$ of the
prefactor of $x^F$ in eq. \eqref{spfin}. This results in
\begin{equation} \label{rrsym}
\begin{Bmatrix} n \\ F \end{Bmatrix}_k =
\sum_{\sum i a_i = F} \prod_{i=1}^{k-1}
\binom{\frac{i n}{k} +\bigl((\II_{k-1}-\CC_{k-1}) \cdot {\bf a}\bigr)_i}{a_i}
\ .
\end{equation}
With this result, we arrive at the following counting formula
for the Read-Rezayi states (for general $k$)
\begin{equation}
\label{rcgk}
\#_{\rm RR} (N,\Delta N_\phi,k) = 
\sum_{F}
\begin{Bmatrix}n\\F\end{Bmatrix}_k
\binom{\frac{N-F}{k} + n}{n} \ ,
\end{equation}
with $n=k \Delta N_\phi$.
To make the above (in particular the symbols $\{ \}_k$)
more explicit, we will discuss the 
$k=2$ (i.e. the pfaffian) and $k=3$ cases. For the pfaffian state
counting, we need to know the symbols $\{ \}_2$.
Eq. \eqref{rrsym} with $k=2$ gives
$\{\begin{smallmatrix}n\\F\\\end{smallmatrix}\}_2=\tbinom{\tfrac{n}{2}}{F}$.
Of course, this is just the result already found in \cite{rere96a}.
Note that our notation is slightly different with respect to the one
used in \cite{rere96a,gure00a}. In our notation, we denote the
number of created quasiholes by $n$. In \cite{rere96a,gure00a},
$n$ denoted the number of extra fluxes, which is denoted by
$\Delta N_\phi$ in this paper.

Although the method described above seems to be unnecessarily
complicated to reproduce this result, it is very useful for
obtaining closed expressions for $k>2$. As an illustration,
we will discuss the case $k=3$, and compare our results with
\cite{gure00a}. For $k=3$, the polynomials are given by 
the following expression
\begin{equation} \label{su2k3pol}
Y_{n}(x;q,3)= \sum_{a,b} q^{\frac{2}{3} (a^2+b^2+ab)} x^{a+2b}
\qbin{\frac{n}{3} - \frac{a+2b}{3}}{a}
\qbin{\frac{2n}{3} - \frac{2a+b}{3}}{b} \ .
\end{equation}
Indeed, these polynomials reduce to the ones in \cite{gure00a},
upon setting $q=1$. The symbols $\{ \}_3$ are now
easily written down 
\begin{equation} \label{su2k3sym}
\begin{Bmatrix}n\\F\end{Bmatrix}_3 = 
\sum_{a+2b=F} 
\binom{\frac{n}{3}-\frac{a+2b}{3}}{a}
\binom{\frac{2n}{3}-\frac{2a+b}{3}}{b} \ .
\end{equation}
Note that only a finite number of terms contribute to the sum in 
eq. \eqref{su2k3sym}. In fact, this is true for all the symbols
\eqref{rrsym} with $n$ finite. 

The fusion rules for the spin field $\sigma$ which is part of the
quasihole operator at level $k=3$ (see \cite{rere99a}), can be
encoded in a Bratteli diagram with the same structure as the
diagram \ref{brat} (note that the fields differ, of course).
This is a consequence of the rank-level duality
$\sut_3 \leftrightarrow \sud_2$ (see \S 16.6 in \cite{frma97a}). 
Thus the total intrinsic degeneracy for the $k=3$ Read-Rezayi state
with $n$ quasiholes is given by $d_n = \cF (n-2)$. Indeed, by summing the
symbols $\{\begin{smallmatrix}n\\F\\\end{smallmatrix}\}_3$ over 
$F$, this result is reproduced. 

The angular momentum multiplets can be found in the same way as
described in section \ref{nassk2}.
Let us note that for $k=1$ the only degeneracy factor remaining
is $\tbinom{N+n}{n}$, which is precisely the orbital factor for
the Laughlin states with quasiholes present. This was of course
to be expected, as the $k=1$ Read-Rezayi states are in fact the
Laughlin states. 

To conclude the discussion on the counting for the Read-Rezayi
states, we would like to mention that the numerical studies
as presented for $k=3$ in \cite{gure00a} are in complete agreement
with the counting formulas. At this point, no numerical
results are available for $k\geq 4$. In the following
section we will turn our attention to the counting of the NASS states
for general level.

\section{Counting formulas for the NASS states}
\label{nasscount}

In this section, we describe the counting for the NASS states
at general level $k$. We will closely follow the procedure of the
previous sections, that is, we  start by writing down the chiral
character corresponding to the $\sud_k/\ue^2$ parafermions
\cite{kuna93a,ge95a}
\begin{equation} \label{chsu3k}
{\rm ch} (x_1,x_2;q,k) =  \sum_{a_i,b_j}
\frac{q^{\frac{1}{2}(\blin{\bf a}{\CC_{k-1}}{\bf a}+
\blin{\bf b}{\CC_{k-1}}{\bf b}-\blin{\bf a}{\CC_{k-1}}{\bf b})}}
{\prod_{i,j} (q)_{a_i} (q)_{b_j}} x_1^{i a_i} x_2^{j b_j} \ ,  
\end{equation}
where we used the same notation as in eq. \eqref{chsu2k}.
This character is of the UCPF form with the K-matrix equal to the
K-matrix of the $\sud_k/\ue^2$ parafermions:
$\KK = \bigl( \begin{smallmatrix} 2 & -1 \\ -1 & 2 \\ \end{smallmatrix}
\bigr) \otimes \A^{-1}_{k-1}$ \cite{arbo01b}.
The recursion relations corresponding to the basis of this
theory can be written in the following way
\begin{eqnarray}
Y_{({\bf l},{\bf m})} (x_1,x_2;q,k) &=&
Y_{({\bf l}-{\bf e}_i,{\bf m})} + x_1^i q^{l_i-\frac{i(k-i)}{k}}
Y_{({\bf l}-\CC_{k-1}\cdot{\bf e}_i,
{\bf m}+\frac{1}{2} \CC_{k-1}\cdot{\bf e}_i)} \nonumber \\
Y_{({\bf l},{\bf m})} (x_1,x_2;q,k) &=&
Y_{({\bf l},{\bf m}-{\bf e}_j)} + x_2^j q^{m_j-\frac{j(k-j)}{k}}
Y_{({\bf l}+\frac{1}{2} \CC_{k-1}\cdot{\bf e}_j,
{\bf m}-\CC_{k-1}\cdot{\bf e}_j)}  \ .
\end{eqnarray}
Once again, we solve the recursion relations by matching these
recursions to eq. \eqref{ucpfrec}. The truncated characters take the form
\begin{align}
Y_{({\bf l},{\bf m})}(x_1,x_2;q,k) =& \sum_{a_i,b_j}
q^{\frac{1}{2}(\blin{\bf a}{\CC_{k-1}}{\bf a}+
\blin{\bf b}{\CC_{k-1}}{\bf b}-\blin{\bf a}{\CC_{k-1}}{\bf b})}
\times \nonumber \\ 
& \prod_{i=1}^{k-1} x_1^{i a_i}
\qbin{\bigl({\bf l}+(\II_{k-1}-\CC_{k-1}) \cdot {\bf a}
+\frac{1}{2} \CC_{k-1} \cdot {\bf b}\bigr)_i}{a_i}
\times \nonumber \\  
& \prod_{j=1}^{k-1} x_2^{j b_j}
\qbin{\bigl({\bf m}+(\II_{k-1}-\CC_{k-1}) \cdot {\bf b}
+\frac{1}{2} \CC_{k-1} \cdot {\bf a}\bigr)_j}{b_j} \ .
\end{align}
We continue by specifying the parameters $l_i$ and $m_j$. We have 
to use the same construction as in the RR case, with the 
difference that we now need the number of spin-up and down 
particles (denoted by $n_\uparrow$ and $n_\downarrow$)
created by the excess flux. Using $l_i=\frac{i n_\uparrow}{k}$
and $m_j=\frac{j n_\downarrow}{k}$ results in
\begin{align} 
Y_{(n_\uparrow,n_\downarrow)}(x_1,x_2;q,k) =& \sum_{a_i,b_j}
q^{\frac{1}{2}(\blin{\bf a}{\CC_{k-1}}{\bf a}+
\blin{\bf b}{\CC_{k-1}}{\bf b}-\blin{\bf a}{\CC_{k-1}}{\bf b})}
\times \nonumber \\
& \prod_{i=1}^{k-1} x_1^{i a_i}
\qbin{\frac{i n_\uparrow}{k}+\bigl((\II_{k-1}-\CC_{k-1}) \cdot {\bf a}
+\frac{1}{2} \CC_{k-1} \cdot {\bf b}\bigr)_i}{a_i}
\times \nonumber \\
& \prod_{j=1}^{k-1} x_2^{j b_j}
\qbin{\frac{j n_\downarrow}{k}+\bigl((\II_{k-1}-\CC_{k-1}) \cdot {\bf b}
+\frac{1}{2} \CC_{k-1} \cdot {\bf a}\bigr)_j}{b_j} \ . \label{ssfin}
\end{align}
From eq. \eqref{ssfin} we obtain the symbols 
$\{ \begin{smallmatrix} n_{\uparrow} & n_{\downarrow} \\
F_{1} & F_{2} \\ \end{smallmatrix} \}_k$
by taking the limit $q \to 1$ of the coefficient of 
$x_1^{F_1} x_2^{F_2}$
\begin{align}
\begin{Bmatrix} n_{\uparrow} & n_{\downarrow} \\
F_{1} & F_{2} \\ \end{Bmatrix}_k =& 
\sum_{\substack{\sum i a_i = F_1 \\ \sum j b_j = F_2}}
\prod_{i=1}^{k-1}
\binom{\frac{i n_\uparrow}{k}+\bigl((\II_{k-1}-\CC_{k-1}) \cdot {\bf a}
+\frac{1}{2} \CC_{k-1} \cdot {\bf b}\bigr)_i}{a_i} 
\times \nonumber \\
& \prod_{j=1}^{k-1} 
\binom{\frac{j n_\uparrow}{k}+\bigl((\II_{k-1}-\CC_{k-1}) \cdot {\bf b}
+\frac{1}{2} \CC_{k-1} \cdot {\bf a}\bigr)_j}{b_j} \ . \label{4sym}
\end{align}
We know have specified all the ingredients of the counting formula
for the NASS states
\begin{equation} \label{ncgk}
\#_{\rm NASS} (N,\Delta N_\phi,k) = 
\sideset{}{'}
\sum_{N_{\uparrow,\downarrow},n_{\uparrow,\downarrow},F_{1,2}}
\begin{Bmatrix}
n_{\uparrow} & n_{\downarrow} \\
F_{1} & F_{2} \\
\end{Bmatrix}_k
\binom{\frac{N_{\uparrow}-F_{1}}{k} + n_{\uparrow}}{n_{\uparrow}}
\binom{\frac{N_{\downarrow}-F_{2}}{k} + n_{\downarrow}}{n_{\downarrow}} \ ,
\end{equation}
where the prime on the sum indicates the constraints
$N_\uparrow+N_\downarrow = N$,
$n_\uparrow+n_\downarrow = 2 k\Delta N_\phi$ and 
$N_\uparrow-N_\downarrow =  n_\downarrow-n_\uparrow$.
The last constraint is a necessary condition for the state to be a
spin-singlet (for more information on the constraints, see
\cite{arre01a}).

The case $k=2$ was explicitly discussed in section \ref{nassk2}. 
For $k=1$ only the orbital degeneracy factors remain, and we obtain
the counting formula for a particular class of Halperin states
\cite{ha83a}. Indeed, for $k=1$, the NASS states reduce to the
spin-singlet Halperin states. As already mentioned in section
\ref{nassk2}, the counting formula \eqref{ncgk} with the symbols
\eqref{4sym} exactly reproduces the results of the diagonalization
studies for $k=2$ \cite{arre01a}. For $k \geq 3$, no numerical results
are available at the moment. 

\section{Discussion}
\label{discussion}

In this paper, we explained in which way parafermions can be used to 
define clustered quantum Hall states. The statistics of these
parafermions is needed to understand the energy spectrum of the 
clustered states in the presence of quasiholes, as obtained from a 
numerical diagonalization study. We obtained explicit formulas for
the symbols $\{ \}_k$ for two classes of clustered states, needed in
the counting formulas. From these formulas, also the spin and
angular momentum multiplet structure of the quasihole degeneracies can
be extracted.

One can say that the parafermion statistics is a crucial part
in the understanding of the ground state properties of a system of interacting
electrons (via an ultra local interaction) on the sphere in the presence
of a magnetic field. In fact, we know of no other way in which this
energy spectrum can be understood. The observation that ground state
properties of a system of interacting electrons needs the knowledge
of parafermion statistics is interesting by itself.

Recently, another class of paired (clustered) spin-singlet states was
proposed \cite{arla01a}. An interesting property of these states is
that the fundamental excitations over these states show a separation
of the spin and charge degrees of freedom. We believe it should be
possible to repeat the present analysis for these newly proposed states,
though it will be more difficult, because the underlying parafermions
are related to $\mathfrak{so}(5)$, a non simply-laced Lie algebra. 
We leave this as a challenge for future work.

{\em Acknowledgments.} The author thanks P. Bouwknegt, R.A.J.~van~Elburg,
N. Read, E. Rezayi and K. Schoutens for discussions. This research is
supported in part by the foundation FOM of the Netherlands.

\providecommand{\bysame}{\leavevmode\hbox to3em{\hrulefill}\thinspace}

\end{document}